# SUPERCONDUCTIVITY WITH ANGULAR DEPENDENT COUPLING: STRIPES, COULOMB REPULSION AND ENHANCED $T_c$

V. V. Moshchalkov and V. A. Ivanov[1]
Laboratorium voor Vaste-Stoffysica en Magnetisme,
Katholieke Universiteit Leuven,
Celestijnenlaan 200D, B-3001 Leuven, BELGIUM

**Abstract**

We have analysed the effect of intrinsic doping inhomogeneity and the presence of stripes in high-$T_c$ superconductors on coupling $\lambda$ by using a simple analytically solvable model with an angular dependent $\lambda(\varphi)$ represented by a square-well form. We have found that the introduction of the Coulomb repulsion $\lambda_C$, increasing the "contrast" $|\lambda| + |\lambda_C|$, or the depth of the angular modulation of $\lambda(\varphi)$, leads to a remarkable enhancement of $T_c$. This effect can be optimized by combining attractive ($\lambda < 0$) and repulsive ($\lambda > 0$) interactions along stripes and perpendicular to them.

High-$T_c$ superconductors (HTSC) have the $CuO_2$ planes as the most essential element of their crystalline structure. Undoped $CuO_2$ planes are in the antiferromagnetic (AF) Mott-insulator state with a perfect square network formed by the oxygen $2p$ and copper $3d$ orbitals. In the insulating tetragonal phase the $CuO_2$ planes seem to be in a pure two-dimensional (2D) regime. Doping of the $CuO_2$ planes results in the insulator-metal transition and the appearance of the $high$-$T_c$ superconductivity. The central problem, the HTSC community is facing now, is the origin of this remarkable transformation from a "trivial" AF insulating to an "exotic" HTSC state. The key to the solution of this problem lies in finding out the answer to the question whether doping occurs homogeneously, thus keeping the 2D character of the doped $CuO_2$ planes intact, or inhomogeneously, when quasi-1D stripes can be formed, thus leading to the spontaneous reduction of the dimensionality from 2D to quasi-1D [1]. In the early days of the $high - T_c$ physics the idea of the persistence of the 2D character of the doped $CuO_2$ layers has taken a dominant position. This idea of "stretching" the highest possible symmetry (2D) even into the metallic HTSC regime looked at first sight very natural but later on turned out to be quite damaging for revealing the 2D-1D crossover induced by the formation of stripes in the $CuO_2$ planes. Recently a convincing experimental evidence has been rapidly emerging proving the reality of the stripe existence in HTSC's. By now stripes have been seen in a variety of experiments, including photoemission spectroscopy [2, 3], neutron scattering [4-6], NMR [7], transport [8] and other techniques. Strong in-plane anisotropy of electrical and heat transport in untwinned $YBa_2Cu_3O_7$ and $YBa_2Cu_4O_8$ single crystals, attributed before to the presence of $CuO$ chains, can also be interpreted in terms of the intrinsic anisotropy of the $CuO_2$ planes themselves, appearing due to the stripe formation [8].

The inverse distance between the stripes, $1/d_s$, first seems to increase with doping (Fig.1) and then saturates, signaling the onset of the charge leakage from stripes into the interstripe areas. The evolution of stripes with doping can be schematically represented as follows (see Fig.2). For low doping the linearly growing stripe momentum $p_s \sim 1/d_s$ (points $a$ and $b$ in Fig.1) implies that the charge concentration in stripes remains more or less constant, while the increase in doping, results in a decreasing stripe separation $d_s$ (see Fig.2a,b) and a growing momentum $p_x$. At point "$c$" (Fig.1) the doping contrast between stripe and interstripe areas starts to be reduced and charges begin to penetrate the interstripe space (Fig.2c). Finally (point "$d$" in Fig.1 and Fig.2d) a 2D homogeneous jellium-like doping is realized with the stripicity being fully suppressed. We assume here, that the evolution of stripicity with doping, shown in Figs.2, reflects the real situation for high- $T_c$'s with the maximum stripicity at the optimum doping and the zero stripicity in the overdoped regime with $T_c \to 0$.

As a very important (and may be even the most important) implication of the stripes scenario is a strong modification of the in-plane angular dependences $\lambda_C(\varphi)$ and $\lambda(\varphi)$ of the electron-electron Coulomb repulsion, $\lambda_C$, and the attractive coupling constant, $\lambda = \rho V$, in the BCS theory. The balance between electron repulsion and attraction defines eventually $T_c$. Therefore, the enigma of the high-$T_c$ problem might be unveiled if the link between the presence of stripes in the $CuO_2$ planes and the strong angular modulation $\lambda_C(\varphi)$, $\lambda(\varphi)$ is established (Figs.2a-c). *We present here a simple analytically solvable model which clearly demonstrates how stripes can strongly enhance $T_c$. The main idea of our model is to take into account the angular dependence of the coupling constant $\lambda(\varphi)$ and the Coulomb repulsion $\lambda_C(\varphi)$.* The formation of stripes makes the Coulomb repulsion in the direction perpendicular to them ($\varphi = \pi/2$) stronger. Remarkably, the very same effect, *repulsion* for ($\varphi = \pi/2$), simultaneously *promotes* the stronger electron-electron *attraction* in the perpendicular direction, *i.e.* along the stripes ($\varphi = 0$). Moreover, the negative values of the gap $\Delta(\varphi = \pi/2)$, typical for the experimentally found $d_{x^2-y^2}$ symmetry [9], could just then reflect a dominant repulsive character of the electron-electron interaction for $\varphi = \pi/2$ and, due to that, a large amplitude of the attractive electron-electron interaction for $\varphi = 0$. High $T_c$ in materials with a very low doping, being compared with low $T_c$ of normal metals with a high homogeneous doping, thus gives an instructive example of how self-organization of charges in the 2D planes into a superlattice of the 1D stripes leads to a peaceful coexistence of a stronger Coulomb repulsion along one direction ($\varphi = \pi/2$) with a higher $T_c$ along the other ($\varphi = 0$).

We start from the qualitative analysis of the modification of the angular dependence $\lambda(\varphi)$ (Fig. 2d) caused by the formation of metallic stripes. For homogeneous jellium like metal (Fig.2a) (zero"stripicity") the $\lambda(\varphi)$ is angular independent. When a weak stripicity is turned on, the electron density of states along the stripes, $g_\parallel(E)$, becomes larger than that perpendicular to them, $g_\perp(E)$. Since Debye screening radius, $r_D = 1/\sqrt{4\pi g(E)}$, is determined by the electron density of states, the $\lambda(\varphi)$ modulation will appear with a stronger Coulomb repulsion $\lambda_C$ in the direction perpendic-ular to stripes ($\varphi = \pi/2$) and a weaker one for $\varphi = 0$ (Fig.2c). At the same time, due to the difference between $g_\parallel(E)$ and $g_\perp(E)$, the BCS attractive coupling will be larger for $\varphi = 0$. Finally, for an "ultimate" or maximum stripicity (Fig.2b) the angles $\varphi$ close to $\pi/2$ are characterized by a net Coulomb repulsion and, respectively, directions close to $\varphi = 0$ provide an attractive electron-electron interaction. The ultimate stripicity is an example of a very unusual strongly correlated state with a Coulomb-gap like ground state for $\varphi = \pi/2$ and superconducting ground state for $\varphi = 0$. It looks like through the presence of stripes in weakly doped 2D systems nature finds a nice solution how to create a hybrid superconductor with a superconducting gap ($\Delta > 0$) along one direction at the expense (or with the help) of having Coulomb gap ($\Delta < 0$) in the perpendicular direction. For simplicity we approximate the real $\lambda(\varphi)$ function with constant values (Fig.2a, b): $\lambda(\varphi) = \lambda_C$ for $\alpha < |\varphi| < \pi/2$ and $\lambda(\varphi) = -\lambda$ for $|\varphi| < \alpha$, respectively. Such a square-well approximation is well known for representing different *energy* dependences of the coupling constant [10], while here we use it to simplify the *angular* dependence of $\lambda$.

In theory of superconductivity the conventional square-well models, possessing the spherical angular symmetry and an energy anisotropy (energy wells for coupling constants) are introduced by defining the characteristic cut-off energies. In spite of their approximate character, these models had revealed the dependence of superconducting properties on the important interactions and the electron density of states, starting from pioneering publications [11-14]. Square well models were successfully used also for the description of the superconducting condensate in organics [15]. The observation of stripes in $high-T_c$ cuprates has put forward the problem of constructing the model with angular dependent coupling strength parameters in real space and without energy dependence of these parameters.

Under these assumptions we obtain the following simplified version of the superconducting gap equation:

$$\Delta(\psi) = -\int_{-\frac{\pi}{2}}^{\frac{\pi}{2}} \frac{d\varphi}{\pi} Q(\psi,\varphi)\Delta(\varphi) \int_0^{\omega_c} d\xi \frac{\tanh\frac{\sqrt{\xi^2+\Delta^2(\varphi)}}{2T}}{\sqrt{\xi^2+\Delta^2(\varphi)}}, \quad (1)$$

where the kernel $Q(\psi,\varphi)$ of this integral equation corresponds to the coupling constant $\lambda = \rho V$ of the conventional BCS theory and the $\omega_c$ is the energy cut-off. We replace the product of the gap function and the kernel by the product of the averages of these quantities, taken separately. The kernel in each angular range is approximated by square wells of half-widths roughly equal to the cut-off angles, thus resulting in $Q_1 = -\lambda, Q_2 = \lambda_C; Q_{12} = Q_{21} = \lambda_C$. If we assume two gaps, $\Delta$ and $\Delta_C$, the angular dependence of the energy gap in Eq.(1) becomes $\Delta(\varphi) = \Delta\vartheta(\alpha^2 - \varphi^2) + \Delta_C[\vartheta(\pi/2 - \varphi)\vartheta(\varphi - \alpha) + \vartheta(\pi/2 + \varphi)\vartheta(-\varphi - \alpha)]$, where $\vartheta$ is the Heaviside function.

The square-well approximation clearly shows how the different coupling constants $\lambda$, $\lambda_C$ and different cut-offs angles $\alpha$, i.e. the size and shape of kernel $Q(\psi, \varphi)$, affect the properties of the system. It also illustrates the dependence of the superconducting properties on the attractive and repulsive interactions and on the electron density of states. Since our aim is to estimate $T_c$ in logarithmic approximation, the Eq.(1) is reduced to

$$\Delta(\psi) = -\int_{-\frac{\pi}{2}}^{\frac{\pi}{2}} \frac{d\varphi}{\pi} Q(\psi, \varphi) \Delta(\varphi) \ln\left[\xi + \sqrt{\xi^2 + \Delta^2(\varphi)}\right]\Big|^{\omega_c} \quad (2)$$

for the cut-off energy parameter $\omega_c$. In our square-well model of stripes (Fig.2) the integral equation (2) with the chosen kernels leads to the following system of algebraic equations for the order parameter near $T_c$:

$$\begin{array}{rcl} (2\alpha Q_1 Z + \pi)\Delta + (\pi - 2\alpha)Q_{12}Z_C\Delta_C & = & 0 \\ 2\alpha Q_{21} Z\Delta + [\pi + (\pi - 2\alpha)Q_2 Z_C]\Delta_C & = & 0 \end{array}. \quad (3)$$

Here the Z-integrals have been calculated in logarithmic approximation: $Z = \ln(2\gamma\omega_c/\pi T_c) = \ln(1.13\omega_c/T_c)$, $Z_C = \ln[(\omega_c + \sqrt{\omega_c^2 + \Delta_C^2})/\Delta_C]$.

The superconducting critical temperature $T_c$ and the corresponding effective coupling constant $\Lambda$ can be derived from the system of equations (3):

$$\begin{vmatrix} 2\alpha\lambda Z - \pi & -(\pi - 2\alpha)\lambda_C Z_C \\ 2\alpha\lambda_C Z & \pi + (\pi - 2\alpha)\lambda_C Z_C \end{vmatrix} = 0. \quad (4)$$

From Eq.(6) it follows that the superconducting critical temperature is $T_c = (2\gamma/\pi)\omega_c \exp\{-1/\Lambda\}$ ($\ln\gamma = C = 0.577$) with an effective coupling constant

$$\Lambda = \frac{2\alpha}{\pi}\left(\lambda + \lambda_C - \frac{\lambda_C}{1 + \frac{\pi-2\alpha}{\pi}\lambda_C \ln\frac{\omega_c + \sqrt{\omega_c^2 + \Delta_C^2}}{\Delta_C}}\right) \quad (5)$$

It is worth noting that this result does not depend on the "anisotropy" of the energy gap $\Delta(\varphi)$, namely the effective coupling constant $\Lambda$ has the same magnitude for the angular dependent $\Delta$ and $\Delta_C$: $\Delta = a\cos^2\varphi$, $\Delta_C = -b\sin^2\varphi$. The $\Delta_C(\varphi)$ dependence is determined by a density of stripes and their specific configuration.

In Eq.(5) the logarithmic terms in denominator play the role of Tolmachev-Bogoliubov-Tyablikov-Anderson-Morel logarithm [11, 12] depressing the electron-electron repulsion (*c.f.*, its role in Mc-Millan parameter). The effective coupling constant $\Lambda$ can be expressed via amplitudes $\Delta$ and $\Delta_C$:

$$\Lambda = \frac{\frac{2\alpha\lambda}{\pi}}{1 + \frac{\pi-2\alpha}{\pi}\lambda_C \frac{\Delta_C}{\Delta} \ln \frac{\omega_c + \sqrt{\omega_c^2 + \Delta_C^2}}{\Delta_C}}.$$

In the BCS model the limits of the energy integration are symmetric with respect to the chemical potential $\mu$: $(\mu - \omega_c, \mu + \omega_c)$ and $\omega_c \sim \omega_D$ (the Debye frequency). This makes it possible to reduce Eq.(1) to Eq.(4) (energies are counted from the chemical potential $\mu$). The latter, however, is not true for the non-retarded electron mechanism of pairing. For example, in the single-band strongly correlated model (*e.g.*,Hubbard) the energy dispersion relation $\xi = (1 - n/2)\varepsilon - \mu$ defines the dependence of the chemical potential $\mu/w = 3n/2 - 1$ ( for rectangular density of electronic states with the energy band of half-width $w$ ) on the carrier concentration $n$ and, as a consequence, the non-symmetric limits of integrations with respect to energy $\xi$, namely $\omega_1 = wn/2$ (lower limit) and $\omega_2 = w(1-n)/2$. In this case one gets $T_c = 2\gamma/\pi\sqrt{n(1-n)}w\exp\{-1/\Lambda\}$,

$$\Lambda = \frac{2\alpha}{\pi}\left(\lambda + \lambda_C - \frac{\lambda_C}{1 + \frac{\pi-2\alpha}{\pi}\lambda_C Z'}\right)$$

with

$$Z' = \ln\left\{\frac{w}{2\Delta_C}\sqrt{\left[n + \sqrt{n^2 + \left(\frac{2\Delta_C}{w}\right)^2}\right]\left[(1-n) + \sqrt{(1-n)^2 + \left(\frac{2\Delta_C}{w}\right)^2}\right]}\right\}.$$

In the limit of $\Delta_C < w$ around a half-filled correlated energy band the $Z'$ is $Z' = \ln[n(1-n)w/\Delta_C]$.

One of the most essential results of introducing the angular dependent coupling (Figs.2, 3) is the enhancement of superconducting critical temperature $T_c$. It turns out that what matters for the effective coupling $\Lambda$ is the sum of the moduli:$|\lambda| + |\lambda_c|$, *i.e.* the depth of the $\lambda(\varphi)$ angular variation defines $\Lambda$. *This depth can be increased by adding the Coulomb repulsion for certain angles* (see Fig.2b). *As a result, the $T_c$ value is enhanced due to the introduced Coulomb repulsion.* It is very instructive to compare the effect of adding $\lambda_C$ to $\lambda$ on $T_c$, considering several typical cases illustrated in Fig.3. Adding extra modulation to a flat $\lambda(\varphi) = const.$ dependence (Fig.3a) leads to a higher $T_c$ for the case with the Coulomb repulsion (Fig.3b), than without it (Fig.3a). Indeed, according to Eq.(5), for typical values of the energy cutoff $\omega_c$ and the Coulomb parameter $\Delta_C$, the third term in brackets in Eq (5) is smaller than the second one and therefore for the same fixed attraction, $\Lambda$ is higher with the Coulomb repulsion $\lambda_C$ (Fig.3b), than without it (Fig.3b).

A very peculiar case is given in Fig.3c: the admixture of the Coulomb repulsion $\lambda_C > 0$ with $\lambda = 0$ which still leads to a finite $T_c$, (Eq.5) although in this case an attractive interaction as such is completely absent for any angle $\varphi$! Nevertheless, the angular variation of $\lambda$ from $\lambda_C > 0$ to $\lambda = 0$ creates apparently a sort of a potential well (see Fig.3) which still makes the existence of superconductivity possible *without any attractive interaction*. Remarkably, the introduction of the Coulomb repulsion for certain angles increases the depth of the angular modulation of the coupling strength and eventually promotes superconductivity. This surprising property highlights the importance of a finite stripicity (Fig.2a - c) for enhancing the superconducting critical temperature in high-$T_c$ cuprates.

The rapidly growing experimental evidence of the presence of highly inhomogeneous doping and formation of stripes in the $CuO_2$ planes, necessitates the elaboration of the theoretical models taking into account the presence of stripes and the 1D character of the charge transport in the high-$T_c$ cuprates. A very enlightening recent experimental observation of the $d_{x^2-y^2}$ character of $\Delta(\varphi)$ [8] is perfectly compatible with the stripes formation (Fig.2b): $d_{x^2-y^2}$ *means two-fold and not four-fold symmetry*. Moreover, positive $\Delta_{x^2-y^2}(\varphi)$ values along the stripes could be interpreted as directions with a maximum attraction while negative $\Delta_{x^2-y^2}(\varphi)$ are indicative for the interstripe Coulomb repulsion for the direction perpendicular to the stripes.

Summarizing, we have considered the effect of the intrinsic doping inhomogeneity and the presence of stripes in high $T_c$'s on the coupling, by using a simple analytically solvable model with an angular dependent $\lambda(\varphi)$, represented by a square-well form. We have found a remarkable effect of enhancing $T_c$ by introducing the Coulomb repulsion for certain angles $\varphi$. The effective coupling $\Lambda$ and $T_c$ turn out to be dependent on $|\lambda|+|\lambda_C|$, thus proving that the most important factor determining $T_c$ is a $\lambda(\varphi)$ "contrast" or a $\lambda(\varphi)$ "modulation depth", which can be maximized by combining attractive ($\lambda < 0$) and repulsive ($\lambda_C > 0$) interactions along the stripes and perpendicular to them, respectively. We think that the maximum angular $\lambda(\varphi)$ contrast is realized in high $T_c$'s in the vicinity of the optimal doping, where addition of $\lambda_C$ helps a lot in increasing $T_c$. In the overdoped regime, due to the doping of the interstripe areas, the added charge suppresses the existing stripicity (Fig.2c,d) and strongly reduces the angular modulation of $\lambda(\varphi)$, which completely disappears around the point $T_c \simeq 0$ where zero stripicity fully restores the 2D character of the $CuO_2$ planes and the jellium-like character of doping at the expense of the full suppression of $T_c$.

The authors would like to thank Yvan Bruynseraede for stimulating discussions. This work is supported by the Flemish FWO and GOA and the Belgian IUAP Programs.

**On leave from :**

[1]*N. S. Kurnakov Institute of the General and Inorganic Chemistry of the Russian Academy of Sciences, Leninskii prospect 31, 117 907 Moscow, RUSSIA*

**Figure captions**

**Fig.1.**
Schematic dependence of the stripe momentum $p_{xs} \sim 1/d_s$, where $d_s$ is the interstripe distance. The points $a - d$ refer also to the four situations, $(a) - (d)$, with a different stripicity shown in Fig.2.

**Fig.2.**
Angular dependence of the coupling (left panel), stripes in the real space (central panel) and the corresponding modifications in the momentum space (right panel) for four different doping levels $(a) - (d)$, corresponding to the points $(a) - (d)$ in Fig.1. Note that the $d_{x^2-y^2}$ symmetry ( $(b)$, central panel ) is perfectly compatible with the presence of the stripes with $\Delta > 0$ (attraction) and $\Delta < 0$ (repulsion) for the directions along and perpendicular to stripes.

**Fig.3.**
Isotropic $(a)$ and different types of the angular dependent coupling, $(b) - (c)$, considered in our model.

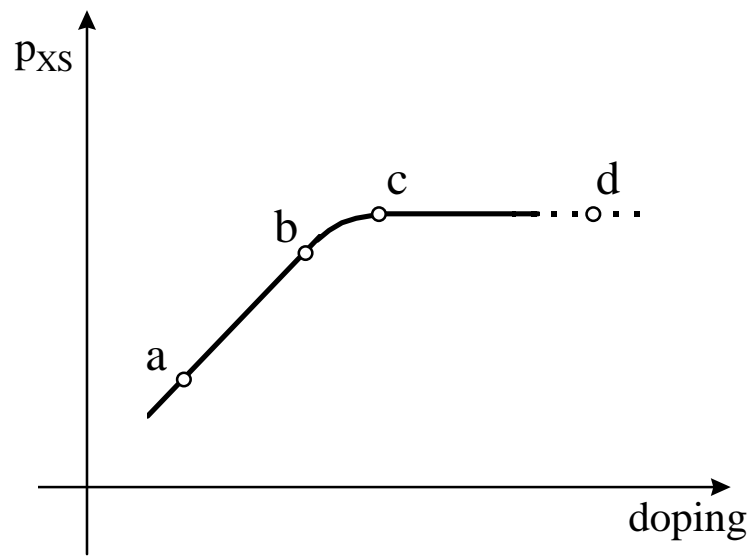

Moshchalkov *et al.*                                                                 Fig.1

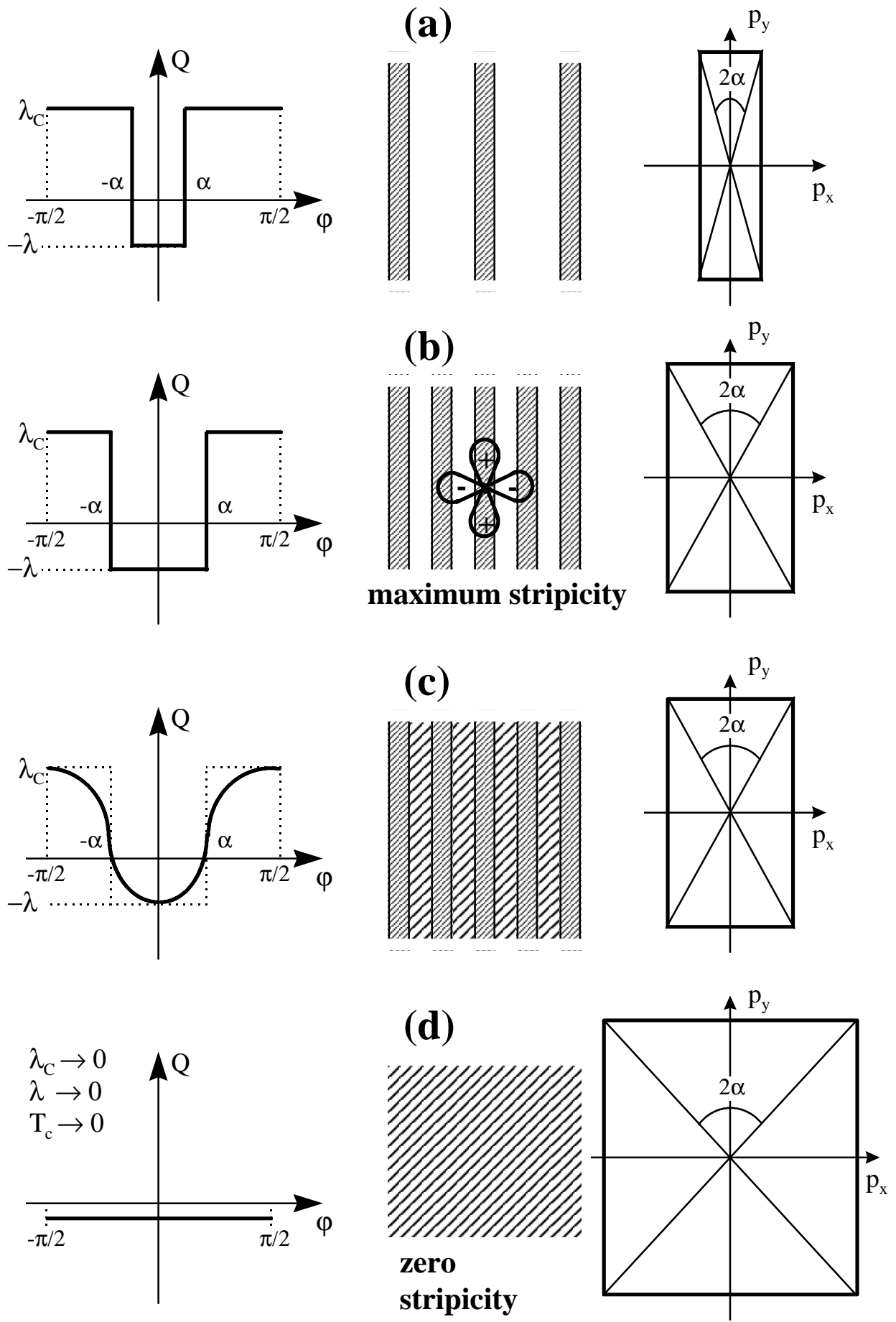

Moshchalkov *et al.*          Fig.2

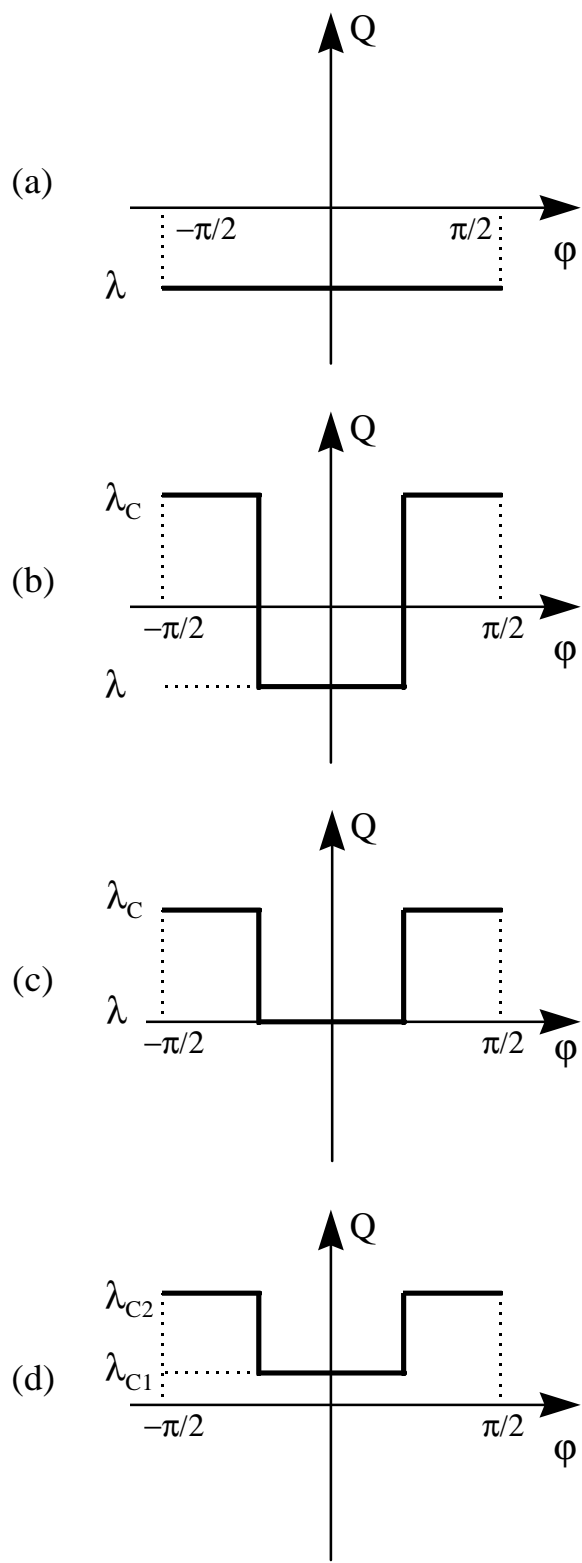

Moshchalkov *et al.*          Fig.3